\documentclass[12pt]{article}  

\usepackage{graphicx,color,epsf} 

\usepackage{amsmath,latexsym}

\begin{document}

\begin{center}
{\LARGE \bf Boundary effects on the thermal stability of a black hole  
	at the centre of a conducting \newline spherical shell
}
\\ 
\vspace{1cm} 
{\large E. S. Moreira Jr.}
\footnote{E-mail: moreira@unifei.edu.br}   
\\ 
\vspace{0.3cm} 
{\em Instituto de Matem\'{a}tica e Computa\c{c}\~{a}o,}  
{\em Universidade Federal de Itajub\'{a},}   \\
{\em Itajub\'a, Minas Gerais 37500-903, Brazil}

\vspace{0.3cm}
{\large July, 2023}
\end{center}
\vspace{0.5cm}


\abstract{ 
This paper reports calculations and analysis of the effects of a 
perfect conducting wall of a very large spherical shell on the stable thermodynamic equilibrium of a black hole sitting at the centre of the shell which is filled with Electromagnetic blackbody radiation.
A parallel is drawn with the case where electromagnetic is replaced by scalar blackbody radiation with Dirichlet or Neumann boundary conditions on the wall. It is found that the value of the shell radius
 above which only blackbody radiation remains in stable
thermodynamic equilibrium 
can be considerably affected by vacuum polarization due to the presence of the wall.
}


\section{Introduction}
\label{introduction}
Five decades after Bekenstein, Bardeen, Carter, Hawking, Hartle,   
Gibbons, Perry \cite{bek73,bar73,haw75,gib76}
and others discovered that a black hole is a thermal object, its associated
quantum degrees of freedom remain unknown. Much understanding has been
gained by considering semi-classical gravity, which consists of treating matter quantum mechanically whereas the background geometry is taken to be classical. Indeed, there are also candidates to  a ``quantum theory of gravity'' (see, for example, ref. \cite{sus05} and references therein) which successfully present explanations for thermal properties of certain special black holes. In spite of a lot of progress, it seems fair to say that there is no consensus
on this matter.

It remains promising to explore the issue in setups where the black hole is in thermodynamic  equilibrium with well known thermal systems.  
One of them is a cavity  with 
the black hole at the centre continuously evaporating by emitting  Hawking radiation and simultaneously accreting blackbody radiation \cite{haw76}. 
Since the pioneering work of Hawking \cite{haw76},
and after early accounts addressing the fundamentals 
of black hole thermodynamics in boxes
\footnote{See also textbooks \cite{nov10,tho17}.}
(e.g., refs. \cite{dav77,hut77,gib78,wil79,pag81}
where heat capacities, phase transitions, the third law of thermodynamics,
Green functions, and
likelihood of black hole formation have been studied),
hundreds of publications followed up covering different aspects, 
which were related, in one way or another, with thermal stability of black holes
in spacetimes of different geometries.
These days the theme is still focus of frequent investigations. Among them are  those in backgrounds where the black hole has a holographic interpretation \cite{sus05}. It should be remarked that such an interpretation is one of the most promising schemes to explaining, consistently (see, e.g., review in ref. \cite{alm21}), the statistical mechanics nature of black holes.

A Schwarzschild black hole of mass $M$
and  electromagnetic blackbody radiation can be 
in stable thermodynamic equilibrium in a cavity only if the temperature $T$ of the system is the 
Hawking temperature \footnote{Fundamental constants have the usual meaning.},
\begin{equation}
T_{\tt H}:=	\frac{\hbar c^3}{8\pi G M\, k},
\label{htemperature}	
\end{equation}
and the cavity's volume $V$ is smaller than a critical value \cite{haw76,nov10,tho17}, 
\begin{equation}
V_{h}=	\frac{{2}^{20}}{5^{4}}\, 3\pi^{2}
\left(\frac{E}{E_{\tt P}}\right)^{5}\ell_{\tt P}^{3},
\label{cvolume}
\end{equation}
where $E$ is the total energy in the cavity, which is given by,
\begin{equation}
	E=Mc^{2}+\frac{\pi^{2}}{15}\frac{(kT)^{4}}{(\hbar c)^{3}}V,
	\label{total-energy}
\end{equation}
and, 
\begin{equation}
	\ell_{{\tt P}}:=\sqrt{\hbar G/c^{3}},\hspace{1.0cm}E_{{\tt P}}:=\sqrt{\hbar c^{5}/G}.	
	\label{planck}	
\end{equation}
In deriving eq. 
(\ref{cvolume}), $E$ 
along with $V$ are taken to be fixed,
whereas $M$ and $T$ are let to evolve such that the entropy,
\begin{equation}
S=\frac{4\pi k}{\hbar c} GM^{2}+\frac{4\pi^{2}}{45}\left(\frac{kT}{\hbar c}\right)^{3}kV,
\label{entropy}
\end{equation}
has a local maximum.

By examining eq. (\ref{total-energy}), one sees that vacuum polarization caused by the non trivial background or the presence of the cavity's wall 
(see, e.g., ref. \cite{dav82})
has been ignored. It can be argued that $V$ is large enough such that vacuum polarization effects would simply yield small corrections 
in eqs. (\ref{total-energy}) and
(\ref{entropy}), and so they perhaps could be neglected.
However, 
small cavities 
should also be examined.
At such a regime, eqs. (\ref{total-energy}) and (\ref{entropy})
do change due to vacuum polarization \cite{bal78},
and thus it remains to answer the question on
the way $V_{h}$ in eq. (\ref{cvolume}) modifies. 
One purpose of this paper is to begin a study of this matter
\footnote{This paper will consider
vacuum polarization caused by the cavity's wall only.}.

Another source of motivation for this work is the following speculation. 
In 1978, Boyer made a surprising discovery that the vacuum energy of a conducting spherical shell tends to expand the shell, instead of shrinking it \cite{boy68}. 
Now, gravitational collapse is a thermodynamic process.
Accordingly, one might wonder that for $V<V_{h}$ the system black hole and blackbody radiation in thermodynamic equilibrium
could be a stage of gravitational collapse of the conducting shell itself.
And, by considering Boyer's discovery mentioned above, it is plausible to conjecture that it could be even a late stage of collapse.
Again, in order to investigate this scenario, 
it would be relevant to determine how vacuum polarization affects 
thermal stability,
in particular how it modifies $V_{h}$ in eq. (\ref{cvolume}). 
The rest of this section consists of a short outline of
calculations and analysis in the following sections.

The literature is rather vague on the nature of the wall that keeps $E$ and $V$ constant in the equations above. It is simply assumed that it is rigid and that no flux of either energy or momentum takes place through it, i.e.,
the cavity's wall is perfectly reflecting.
Nevertheless, any wall is a physical object, even when it is idealized. Thus the proper way of dealing with the problem in the context of quantum fields at finite temperature is to use suitable boundary conditions (see, e.g., the classic paper ref. \cite{bro69}, and a related thermodynamic account in ref. \cite{mor23}).

Consider a perfect conducting spherical shell of radius $R$,
carrying electromagnetic radiation at temperature $T$. As Balian and Duplantier have shown  in ref. \cite{bal78}, at the regime of high temperatures and/or large shells,
i.e., 
\begin{equation}
\frac{kTR}{\hbar c}\gg 1,	
	\label{ht}	
\end{equation}
the Helmholtz free energy in the shell is given by
\footnote{It should be remarked that corrections to the Planckian free energy, which is the term containing $T^{4}V$ in eq. (\ref{fenergy}),
depends very much on the geometry of the cavity and on the type of boundary condition on its walls.
For example, in the case of a perfect conducting cubic cavity, the leading correction is proportional to $T^{2}V^{1/3}$ \cite{amb83}, and thus contrasting considerably with that in eq. (\ref{fenergy})
for a perfect conducting spherical shell, and also with those in eq. (\ref{sfenergy}) for scalar radiation
in a reflecting spherical shell.},
\begin{equation}
{\cal F}=-\frac{\pi^{2}}{45}\frac{(kT)^{4}}{(\hbar c)^{3}}V
-\frac{kT}{4}\left[
{\rm ln}\left(\frac{kTR}{\hbar c}\right)
+
0.769
\right]+\cdots,	
\label{fenergy}	
\end{equation}
where 
\begin{equation}
V=\frac{4}{3}\pi R^{3}.	
\label{volume}	
\end{equation}
It follows from eq. (\ref{fenergy}) the internal energy and entropy,
\begin{equation}
	{\cal U}=\frac{\pi^{2}}{15}\frac{(kT)^{4}}{(\hbar c)^{3}}V
	+\frac{kT}{4}+\cdots,
	\hspace{0.8cm}
	{\cal S}=\frac{4\pi^{2}}{45}\left(\frac{kT}{\hbar c}\right)^{3}kV
	+\frac{k}{4}\left[
	{\rm ln}\left(\frac{kTR}{\hbar c}\right)
	+
	1.769
	\right]+\cdots,	
	\label{venergy}	
\end{equation}
each containing corrections to the familiar Planckian expressions
due to the presence of the spherical shell's perfect conducting wall. A couple of remarks are in order regarding 
eq. (\ref{venergy}) as they will be used later on.
The term $kT/4$ in ${\cal U}$
is often  identified as been a ``classical'' correction, since it does not carry $\hbar$. However, when one sets $\hbar\rightarrow 0$, the Planckian internal energy becomes ``infinite'' and thus $kT/4$ ``vanishes'':
$kT/4$ is simply the $\hbar^{0}$-term of an expansion in powers of $\hbar$. 
Another remark is that the presence of the perfect conducting wall increases the heat capacity of the hot radiation in the shell. Consequently, the blackbody radiation temperature becomes less ``agile'' to change when ${\cal U}$ changes.

Now a black hole 
is set at the shell's centre and let to get in stable thermodynamic equilibrium with the hot radiation surrounding  it. Then the vacuum polarization corrections in eq. (\ref{venergy}) are taken into account in eqs. (\ref{total-energy}) and (\ref{entropy}), and the problem of finding a local maximum for $S$ is redone, leading to a new $V_{h}$ [see eq. (\ref{cvolume})]. Such a program is implemented in Sec. \ref{vector}.

In Sec. \ref{scalar}, the problem is reconsidered for spherical shells containing hot scalar radiation, with Dirichlet and Neumann walls. Sec. \ref{conclusion} presents a summary and addresses further issues considered in previous sections, such as  comparison 
between vacuum polarization modifications of eq. (\ref{cvolume})
regarding the nature of the radiation in the spherical shell (electromagnetic or scalar) as well as the type of boundary conditions on its wall (Dirichlet or Neumann).
Sec. \ref{conclusion} ends after  pointing out an extension of the work.
An appendix outlines an alternative derivation of $V_{h}$, closing the paper.


\section{Electromagnetic radiation}
\label{vector}
As has been sketched in Sec. \ref{introduction},
eqs. (\ref{total-energy}) and (\ref{entropy}) are modified by considering eq. (\ref{venergy}), i.e.,
\begin{eqnarray}
&&E=Mc^{2}+\frac{\pi^{2}}{15}\frac{(kT)^{4}}{(\hbar c)^{3}}V
+\frac{kT}{4}+\cdots,	
\label{E}	
\\	
&&S=\frac{4\pi k}{\hbar c} GM^{2}+
\frac{4\pi^{2}}{45}\left(\frac{kT}{\hbar c}\right)^{3}kV
+\frac{k}{4}\left[
{\rm ln}\left(\frac{kTR}{\hbar c}\right)
+
1.769
\right]+\cdots.
\label{S}	
\end{eqnarray}
Ellipses, denoting smaller corrections
[note eq. (\ref{ht})], will be omitted from now on.
It is convenient to defined the dimensionless quantities,
\begin{equation}
m:=\frac{Mc^{2}}{E}, \hspace{1.0cm}
t:=\frac{kT}{4E},
\label{parameters}
\end{equation}
and to recast eq. (\ref{E}) as,
\begin{equation}
m=1-\frac{2^{10}\pi^{3}}{45}\left(\frac{ER}{\hbar c}\right)^{3}t^{4}-t,	
\label{m}	
\end{equation}
where eq. (\ref{volume}) has been used.
In the literature (see, e.g., ref. \cite{tho17}), where only Planckian contributions are considered [cf. eqs. (\ref{total-energy}) and (\ref{entropy})], it is $m$ in eq. (\ref{parameters})
that plays the role of
independent variable. In the present case, where corrections to the Planckian contributions are also taken into account, $t$ in eq. (\ref{parameters})
is the suitable independent variable in the  problem. It follows that $m$ in eq. (\ref{m}) is a function of $t$, for given $E$ and $R$.
By defining another dimensionless quantity, namely
\begin{equation}
s:=\frac{\hbar c^{5}S}{4\pi kGE^{2}},	
\label{ps}
\end{equation}
eq. (\ref{S})  leads to,
\begin{equation}
s=m^{2}+
\frac{2^{8}\pi^{2}}{135}\frac{ER^{3}c^{2}}{\hbar^{2}G}t^{3}
+\frac{\hbar c^{5}}{16\pi GE^{2}}\left[
{\rm ln}\left(\frac{4ER}{\hbar c}t\right)
+
1.769
\right],
\label{s}
\end{equation}
which is also a function of $t$. At this point, it should be remarked that, since $ER\gg kTR \gg \hbar c$ [see eqs. (\ref{ht}) and (\ref{E})], $t$ in eq. (\ref{parameters}) is small, 
but not that much: 
\begin{equation}
\frac{\hbar c}{ER} \ll t \ll 1.	
\label{tbounds}
\end{equation}	

A comment on the shell's radius is also worth making at this point. As shall be seen soon, a necessary condition for thermodynamic equilibrium is still that $T=T_{\tt H}$ [see eq. (\ref{htemperature})], as expected. Now, noticing eq. (\ref{ht}),
it follows that,
\begin{equation}
R\gg R_{\tt S},	
\label{lradius}
\end{equation}
with $R_{\tt S}$ denoting the Schwarzschild radius of the black hole in the shell:
\begin{equation}
 R_{{\tt S}}:=\frac{2GM}{c^{2}}.	
\label{sradius}
\end{equation}
That is, the size of the shell must be large compared with that of  the hole. If $R$  were such that the inequality in eq. (\ref{ht})
were reversed, then the regime would be that of low temperatures and/or small shells, resulting that  eq. (\ref{venergy}) would no longer hold \cite{bal78}. In particular, the dominant contribution in the expression for ${\cal U}$ would be a vacuum energy, instead of the familiar Planckian internal energy. 

Noticing eq. (\ref{m}), derivative of eq. (\ref{s}) with respect to $t$ yields,
\begin{equation}
\frac{ds}{dt}=
\frac{\hbar c^{5}}{16\pi GE^{2}t}\left[
\frac{4^{6}\pi^{3}}{45}\left(\frac{ERt}{\hbar c}\right)^{3}+1\right]
(1-\tau),
\label{ds/dt}
\end{equation}
where,
\begin{equation}
\tau:=32\pi \frac{GE^{2}}{\hbar c^{5}} mt.
\label{tau}
\end{equation}
It follows immediately from eqs. (\ref{htemperature}), (\ref{parameters}) and (\ref{tau}) that,
\begin{equation}
	\tau=\frac{T}{T_{\tt H}}.
\label{tau2}	
\end{equation}
Clearly, blackbody radiation and the black hole cannot coexist together in thermodynamic equilibrium if the maximum value of $\tau$ is less than unity. Otherwise, according to eq. (\ref{ds/dt}),
$s$ would grow for all $t$ and the black hole would evaporate leaving only blackbody radiation behind.

The maximum value of $\tau$ happens when $d\tau/dt=0$, i.e.[see eqs. (\ref{m}) and (\ref{tau})],
\begin{equation}
1-\frac{2^{10}\pi^{3}}{9}\left(\frac{ER}{\hbar c}\right)^{3}t^{4}-2t=0.
\label{ttau}
\end{equation}
By recalling eq. (\ref{tbounds}), the solution of eq. (\ref{ttau}) is given by \footnote{As mentioned previously, keeping main contributions only.
}:
\begin{equation}
t_{\tau}=\left(\frac{9}{2^{10}\pi^{3}}\right)^{1/4}
\left(\frac{\hbar c}{ER}\right)^{3/4}
\left[1-\frac{1}{4}\left(\frac{9}{(4\pi)^{3}}\right)^{1/4}
\left(\frac{\hbar c}{ER}\right)^{3/4}\right],	
	\label{ttau2}
\end{equation}
corresponding to,
\begin{equation}
m(t_{\tau})=\frac{4}{5}	
-\frac{3}{10}
\left(\frac{9}{(4\pi)^{3}}\right)^{1/4}
\left(\frac{\hbar c}{ER}\right)^{3/4},
\label{mtau}
\end{equation}
where eq. (\ref{m}) has been used.
Now, eqs. (\ref{tau}), (\ref{ttau2}) and (\ref{mtau}) yield,
\begin{equation}
\tau(t_{\tau})=\frac{2^{6}\pi}{5}\frac{GE^{2}}{\hbar c^{5}}
\left(\frac{9}{(4\pi)^{3}}\right)^{1/4}
\left(\frac{\hbar c}{ER}\right)^{3/4}
\left[1-\frac{5}{8}\left(\frac{9}{(4\pi)^{3}}\right)^{1/4}
\left(\frac{\hbar c}{ER}\right)^{3/4}\right],	
\label{maxtau}
\end{equation}
as the maximum value of $\tau$.

By taking into account the comments just after eq. (\ref{tau2})
and noting that $\tau(t_{\tau})$ 
in eq. (\ref{maxtau}) diminishes as $R$ increases with $E$ fixed,
one can determine the shell's critical volume $V_{h}$ by solving 
$\tau(t_{\tau})=1$ for $R$ [see eqs. (\ref{planck}) and (\ref{volume})], obtaining: 
\begin{equation}
V_{h}=\frac{{2}^{20}}{5^{4}}3\pi^{2}
\left(\frac{E}{E_{\tt P}}\right)^{5}\ell_{\tt P}^{3}-
\frac{{2}^{13}}{5^{2}}3\pi
\left(\frac{E}{E_{\tt P}}\right)^{3}\ell_{\tt P}^{3},	
\label{vvolume}
\end{equation}
which should be compared with eq. (\ref{cvolume}).
An interesting aspect of eq. (\ref{vvolume}) is that the second term on its rhs is ``classical'' as it is $kT/4$ in eq. (\ref{E}) [see remark in the text just after eq. (\ref{venergy})].
Indeed, 
$$
2\frac{E}{E_{\tt P}}\ell_{\tt P}
=\frac{2GE}{c^{4}}.
$$
Thus, the vacuum polarization effect of a perfect conducting wall is to reduce the value in eq. (\ref{cvolume}) by nearly a  hundred of 
times the volume of a Schwarzschild black hole of mass $E/c^{2}$ [see eq. (\ref{vvolume})]. 
It should be pointed out that,
since $R_{{\tt S}}>l_{{\tt P}}$,
eqs. (\ref{lradius}) and (\ref{vvolume}) imply that,
\begin{equation}
	\frac{E}{E_{{\tt P}}}\gg 1.	
	\label{lenergy}	
\end{equation}
In order to have a crude idea of the size of each term in eq. (\ref{vvolume}), one sets $E/E_{\tt P}\approx 10$ and $E/E_{\tt P}\approx 1$
to see that the reduction is about $0.06\%$ and $6\%$
of the main contribution, respectively. As it will be shown in the next section,
the corresponding correction for scalar radiation is much larger.

\section{A parallel: scalar radiation}
\label{scalar}
At this section, electromagnetic is replaced by scalar 
blackbody radiation coexisting with the Schwarzschild  black hole in a spherical shell of radius $R$.
Dirichlet and Neumann boundary conditions on the shell's wall will be addressed. The results will be compared with those in the previous section.

Still considering the regime of high temperatures and/or large shells [see eq. (\ref{ht})],
the Helmholtz free energy for the scalar radiation 
can be obtained from Dowker's general formula in ref.
\cite{dow84}, and it is given by [cf. eq. (\ref{fenergy})]
\footnote{Recall that ellipses are being omitted.}
\begin{equation}
{\cal F}=-\frac{\pi^{2}}{90}\frac{(kT)^{4}}{(\hbar c)^{3}}V
\pm\frac{\zeta(3)}{8\pi}\frac{(kT)^{3}}{(\hbar c)^{2}}A,
\label{sfenergy}	
\end{equation}
upper sign for Dirichlet, lower sign for Neumann
\footnote{As in the whole text.}, and where $A=4\pi R^{2}$
[note also eq. (\ref{volume})]. Then eq. (\ref{venergy})
gives place to,
\begin{eqnarray}
	{\cal U}=\frac{\pi^{2}}{30}\frac{(kT)^{4}}{(\hbar c)^{3}}V
	\mp\frac{\zeta(3)}{4\pi}\frac{(kT)^{3}}{(\hbar c)^{2}}A,
&&	
	{\cal S}=\frac{2\pi^{2}}{45}\left(\frac{kT}{\hbar c}\right)^{3}kV
	\mp\frac{3\zeta(3)}{8\pi}\left(\frac{kT}{\hbar c}\right)^{2}kA.
	\label{senergy}	
\end{eqnarray}
Unlike the correction to the Planckian internal energy in eq. (\ref{venergy}), ${\cal U}$ in eq. (\ref{senergy}) carries  ``quantum'' corrections. Notice also that the Dirichlet wall makes  the scalar radiation temperature more ``agile'' to change when 
${\cal U}$ varies, whereas the effect caused by the Neumann wall is the opposite one \footnote{Like that for electromagnetic radiation. See Sec. \ref{vector}.}.

Correspondingly, eqs. (\ref{E}) and (\ref{S}) are replaced by their scalar counterparts:
\begin{eqnarray}
	&&E=Mc^{2}+\frac{\pi^{2}}{30}\frac{(kT)^{4}}{(\hbar c)^{3}}V
	\mp\frac{\zeta(3)}{4\pi}\frac{(kT)^{3}}{(\hbar c)^{2}}A,
	\nonumber	
	\\	
	&&S=\frac{4\pi k}{\hbar c} GM^{2}+\frac{2\pi^{2}}{45}\left(\frac{kT}{\hbar c}\right)^{3}kV
	\mp\frac{3\zeta(3)}{8\pi}\left(\frac{kT}{\hbar c}\right)^{2}kA,
	\nonumber
\end{eqnarray}
leading to [cf. eqs. (\ref{m}) and (\ref{s})],
\begin{eqnarray}
&&	
m=1-\frac{2^{9}\pi^{3}}{45}\left(\frac{ER}{\hbar c}\right)^{3}t^{4}
\pm \zeta(3)2^{6}\left(\frac{ER}{\hbar c}\right)^{2}t^{3},
\label{sm}		
\\
&&	
s=m^{2}+
\frac{2^{7}\pi^{2}}{135}\frac{ER^{3}c^{2}}{\hbar^{2}G}t^{3}
\mp \frac{\zeta(3)6}{\pi}\frac{R^{2}c^{3}}{\hbar G}t^{2},	
\label{ss}		
\end{eqnarray}
where eqs. (\ref{parameters}) and (\ref{ps}) have been used again.
It should be remarked that eqs. (\ref{tbounds}) and (\ref{lradius})
still hold. Now, proceeding as in the previous section, eqs. (\ref{sm})
and  (\ref{ss}) lead to [cf. eq. (\ref{ds/dt})]:
\begin{equation}
	\frac{ds}{dt}=
	\frac{\hbar c^{5}}{32\pi GE^{2}t}\left[
	\frac{4^{6}\pi^{3}}{45}\left(\frac{ERt}{\hbar c}\right)^{3}\mp\zeta(3)2^{7}3\left(\frac{ERt}{\hbar c}\right)^{2}\right]
	(1-\tau),
	\label{sds/dt}
\end{equation}
where $\tau$ is defined in eq. (\ref{tau}) and it satisfies eq. (\ref{tau2}).
Again, by examining eq. (\ref{sds/dt}), one sees that 
for a black hole and blackbody radiation to coexist in the shell,
the maximum value of $\tau$ must not be less than
unity. Then, noticing eqs. (\ref{sm}) and (\ref{tau}), 
the maximum value of $\tau$ happens when $d\tau/dt=0$ as before,
resulting that eq. (\ref{ttau}) gives place to,
$$
	1-\frac{2^{9}\pi^{3}}{9}\left(\frac{ER}{\hbar c}\right)^{3}t^{4}\pm\zeta(3)2^{8}\left(\frac{ER}{\hbar c}\right)^{2}t^{3}=0,
$$
whose solution is given by [cf. eq. (\ref{ttau2})]:
\begin{equation}
	t_{\tau}=\left(\frac{9}{2^{9}\pi^{3}}\right)^{1/4}
	\left(\frac{\hbar c}{ER}\right)^{3/4}
	\left[1\pm\zeta(3)\left(\frac{9}{2\pi^{3}}\right)^{3/4}
	\left(\frac{\hbar c}{ER}\right)^{1/4}\right].	
	\label{stau2}
\end{equation}
Now, by setting $t=t_{\tau}$ in eq. (\ref{sm}), it results that,
  \begin{equation}
  	m(t_{\tau})=\frac{4}{5}	
  	\pm \frac{\zeta(3)}{5}
  	\left(\frac{9}{2\pi^{3}}\right)^{3/4}
  	\left(\frac{\hbar c}{ER}\right)^{1/4},
  	\label{smtau}
  \end{equation}
which corresponds to eq. (\ref{mtau}).
Then, the maximum value of $\tau$ is given by,
$$
	\tau(t_{\tau})=\frac{2^{7}\pi}{5}\frac{GE^{2}}{\hbar c^{5}}
	\left(\frac{9}{2^{9}\pi^{3}}\right)^{1/4}
	\left(\frac{\hbar c}{ER}\right)^{3/4}
	\left[1\pm\zeta(3)\, 5\, 2^{4}\left(\frac{9}{2^{9}\pi^{3}}\right)^{3/4}
	\left(\frac{\hbar c}{ER}\right)^{1/4}\right],	
$$
where eqs. (\ref{tau}), (\ref{stau2}) and (\ref{smtau}) 
have been used.
Finally, considering the arguments just before eq. (\ref{vvolume}), one ends up with,
\begin{equation}
	V_{h}=\frac{{2}^{21}}{5^{4}}3\pi^{2}
	\left(\frac{E}{E_{\tt P}}\right)^{5}\ell_{\tt P}^{3}
	\pm\zeta(3)\left[
	\frac{{2}^{56}}{5^{8}}\frac{3^{7}}{\pi}
	\left(\frac{E}{E_{\tt P}}\right)^{13}\right]^{1/3}
	\ell_{\tt P}^{3},	
	\label{svolume}
\end{equation}
which is the scalar version of eq. (\ref{vvolume}), for Dirichlet and Neumann boundary conditions on the shell's wall. Unlike the correction in eq. (\ref{vvolume}) due to the presence of a conducting wall,
those corrections in eq. (\ref{svolume}) carry $\hbar$ 
\footnote{See comments just after eq. (\ref{vvolume}).}.

Considering  eq. (\ref{lenergy}), which also holds for scalar radiation, it should be remarked that 
by setting $E/E_{\tt P}\approx 10$ and $E/E_{\tt P}\approx 1$
in eq. (\ref{svolume}),
the corrections are about $13\%$ and $60\%$
of the main contribution, respectively. 
As has been previously mentioned
\footnote{See text closing Sec. \ref{vector}.}, 
these are far larger than 
the corresponding correction in eq. (\ref{vvolume}).

\section{Final remarks}
\label{conclusion}

A Black hole is a thermal object, and it can be in stable thermodynamic equilibrium with blackbody radiation, if the box
where they coexist is not that large. This paper reported a study of the effects on thermal stability of acknowledging vacuum polarization caused by the wall of a spherical shell which contains a Schwarzschild black hole at its centre and it is filled with either electromagnetic or scalar hot radiations.
The critical volume $V_{h}\propto (E/E_{\tt P})^{5}\ell_{\tt P}^{3}$ 
above which the two systems cannot coexist in stable thermodynamic equilibrium is well known. It was shown here that this value is reduced by a term proportional to $(E/E_{\tt P})^{3}\ell_{\tt P}^{3}$  
in the case of electromagnetic radiation in a perfect conducting shell, and it is increased
by a term proportional to $\pm(E/E_{\tt P})^{13/3}\ell_{\tt P}^{3}$ 
in the case of scalar radiation in a shell with Dirichlet ($+$ sign) or Neumann ($-$ sign) walls.

As is well known, a Schwarzschild black hole has a negative heat capacity, $C<0$ [see eq. (\ref{c})]. It can be in stable thermodynamic equilibrium with blackbody radiation, with heat capacity $C_{V}>0$, only if, 
\begin{equation}
C_{V}<|C|,	
	\label{hc}
\end{equation}
as a basic calculation in thermodynamics may show
\footnote{The inequality in eq. (\ref{hc}) is equivalent to the requirement that total entropy has a local maximum at the equilibrium temperature.}. That is, eq. (\ref{hc}) says that the blackbody radiation must be more ``agile'' than the black hole to catch up the temperature of the latter when it rushes away from equilibrium. by replacing ``$<$'' in eq. (\ref{hc})
by ``$=$'', one obtains $V_{h}$ in eqs. (\ref{vvolume}) and (\ref{svolume})
(see Appendix \ref{a1}).

A natural extension of the present work would be an investigation when  the inequality in eq. (\ref{ht}) is less stringent, 
such that low temperatures and/or small shells could be considered.
In an early work, Pav\'{o}n and Israel have speculated that the first term in the rhs of eq. (\ref{svolume}) is a good approximation for $V_{h}$ ``even for Planck-mass black holes and for box radii comparable with the size of the black hole'' \cite{pav84}.
Recalling  that by setting $E/E_{\tt P}\approx 1$ 
in eq. (\ref{svolume})
the corrections reach $60\%$
of the main contribution, it seems that the conclusion in Ref. \cite{pav84} needs to be re-examined. 
At such a regime, 
vacuum polarization due to non trivial black hole geometry will most likely have to be taken into account, as has indeed been done in Ref. \cite{pav84}. Nevertheless, vacuum polarization due to the wall, which was the focus here,
certainly cannot be neglected and,
perhaps, vacuum energy may play a crucial role.

\appendix 
\section{$V_{h}$ following from  eq. (\ref{hc})}
The material below makes the text rather self-contained
by showing that eq. (\ref{vvolume}) follows from eq. (\ref{hc}). Analogous calculations 
can be used to derive eq. (\ref{svolume}) from  eq. (\ref{hc}) also.
\label{a1}

The heat capacity $C_{V}=\partial_{T}{\cal U}$ of electromagnetic radiation in a perfect conducting spherical shell follows from eq. (\ref{venergy}) and it is given by,
 \begin{equation}
C_{V}=\frac{4\pi^{2}}{15}k\left(\frac{kT}{\hbar c}\right)^{3}V
	+\frac{k}{4},
\label{cv}
\end{equation}
where only main contributions have been kept.
Noting eq. (\ref{htemperature}), one sets
$T=T_{\tt{H}}$ in eq. (\ref{cv}), resulting that,
\begin{equation}
C_{V}=\frac{k}{2^{7}15\pi}V\left(\frac{c^{2}}{MG}\right)^{3}
+\frac{k}{4}.
\label{cv2}
\end{equation}
The heat capacity $C=\partial_{T_{\tt{H}}}(Mc^{2})$
of the Schwarzschild black hole at the centre of the shell is:
\begin{equation}
C=-8\pi k\frac{GM^{2}}{\hbar c}.
\label{c}
\end{equation}
Now, by considering the comments in the paragraph of eq. (\ref{hc}),
one sets $C_{V}=|C|$ and solves for $V$, finding that,
\begin{equation}
V_{M}=2^{10}15\pi^{2}\frac{G^{4}M^{5}}{\hbar c^{7}}-2^{5}15\pi \frac{G^{3}M^{3}}{c^{6}},
\label{Vm}
\end{equation}
where eqs. (\ref{cv2}) and (\ref{c})
have been used.

The next step is to replace in eq. (\ref{E}), $T=T_{\tt{H}}$ and $V=V_{M}$,
yielding,
\begin{equation}
E=\frac{5}{4}Mc^{2}+\frac{3}{2^{7}\pi}\frac{\hbar c^{3}}{GM}.
\label{M}
\end{equation}
As usual, only the leading correction to $5Mc^{2}/4$ has been considered in eq. (\ref{M}).
The last step is to use eq. (\ref{M}) in eq. (\ref{Vm}), which leads to eq. (\ref{vvolume}), again.

\vspace{1cm}
\noindent{\bf Acknowledgements} -- 
Work partially supported by
``Funda\c{c}\~{a}o de Amparo \`{a} Pesquisa do Estado de Minas Gerais'' (FAPEMIG)
and by ``Coordena\c{c}\~{a}o de Aperfei\c{c}oamento de Pessoal de N\'{\i}vel Superior'' (CAPES).

\end{document}